# A Screen-printed, Silver-nanowire-based Optically Transparent Wideband Reconfigurable Intelligent Surface for 5G mmWave Applications

Yiming Yang, *Student Member, IEEE*, Mohammad Vaseem, *Member, IEEE*, Ruiqi Wang, Behrooz Makki, *Senior Member, IEEE*, and Atif Shamim, *Fellow, IEEE*

*Abstract*—The reconfigurable intelligent surface (RIS) can help establish secondary line-of-sight links for 5G mmWave and beyond applications to cater for the large path loss and blockage in millimeter-wave direct links. It is desirable that the RISs are optically transparent so that they can be placed on building windows, etc. However, traditional transparent designs with indium tin oxide films have challenges of high loss and low transparency, while metallic meshes suffer from a large feature size for millimeter wave designs. In this letter, we propose an optically transparent RIS using screen-printed silver nanowires (AgNW) and PIN diodes. The AgNW ink has more than 80% optical transparency and a decent conductivity of $5 \times 10^6$ S/m. The RIS array size is 9x20, and each column serially connects 9 unit cells. The RIS can work from 23.5 GHz to 29.5 GHz with 30° angular stability. The measurement results of the RIS show up to 15 dB gain enhancement within ±30° angle range, while maintaining high optical transparency and large bandwidth.

## I. INTRODUCTION

There have been plenty of examples of reconfigurable intelligent surfaces (RIS) in recent literature where they help establish the secondary line-of-sight for 5G mmWave and 6G communications [1], [2], [3]. However, the implementations on building windows, solar cell panels, automotive windshields, and many other scenarios, require optical transparency, with which most of the common Printed Circuit Board (PCB) based RISs are not compatible. For an optically transparent RIS, the opaque PCB must be replaced with an optically transparent substrate, and more importantly the conductor must also be optically transparent. This enables the light transmittance through the RIS structure while maintaining its capability to manipulate the impinged electromagnetic waves.

Numerous examples have proposed optically transparent RIS, which use common techniques such as lithographed metallic meshes, thin films such as indium-tin-oxide (ITO) films, and conductive polymers [4], [5]. The metallic meshes have been demonstrated for sub-10 GHz applications with acceptable transparency [6], [7]. However, as the frequency increases to millimeters, the feature size is too small for metallic meshes to function properly. The ITO film as an alternative can better balance the RF loss with optical transparency. However, the conductivity is still below $10^6$ S/m, and the fabrication cost is high due to the high price of indium, and the expensive deposition process [8], [9]. Conductive polymers [10], [11] have lower conductivity than thin films and metallic meshes, and are thus not suitable for millimeter-wave applications.

In this letter, we develop and test an optically transparent RIS using screen-printed silver nanowires (AgNW) and PIN diodes. To solve the issues of RF loss and low transparency, conductive nanowires such as silver nanowires and copper nanowires offer the best trade off [4], [5]. To increase the manufacturability and enable large-scale applications, we use a custom silver-nano wire ink [5], [12] with screen-printing process [13]. The AgNW ink consists of silver nano-wires mixed with a polymer, where the diameter of the wires is tens of nanometers and the length is a few microns. After sintering, the nanowires can create meshes with large opening area, allowing light to pass through. Since the nanowires have much smaller feature sizes than the metallic meshes, the printed AgNW can be more suitable for 5G mmWave and 6G bands. Further, to fabricate the conductor pattern, we put our design on a stencil and use screen-printing. The stencil selectively allows the ink onto the substrate based on the pattern of the screen, which is a much more efficient method (faster and lower cost) than PCB manufacturing due to its additive manufacturing nature.

In following sections, we begin with the unit cell design and analysis of angular stability and phase-shift mechanism. We then propose and prototype the serially connected RIS columns with biasing circuitry. Finally, we carry out a field measurement of the gain enhancement with different reflection angles.

This work was supported by Ericsson Research under Grant OSR #4606 (*Corresponding author: Yiming Yang*).

Yiming Yang, Mohammad Vaseem, Ruiqi Wang, and Atif Shamim are with the Computer, Electrical and Mathematical Sciences and Engineering Division, King Abdullah University of Science and Technology (KAUST), Thuwal 23955, Saudi Arabia (e-mail: yiming.yang@kaust.edu.sa; mohammad.vaseem@kaust.edu.sa; ruiqi.wang.1@kaust.edu.sa; atif.shamim@kaust.edu.sa).

Behrooz Makki is with Ericsson Research, 417 56 Gothenburg, Sweden (e-mail: behrooz.makki@ericsson.com)

Color versions of one or more of the figures in this article are available online at http://ieeexplore.ieee.org

## II. RIS Unit Cell and Array Design

The layer stack up and the top view of the proposed RIS unit cell are shown in Fig. 1. From the stack up in Fig. 1 (a), we can see that the unit cell consists of a glass substrate and two AgNW conductor layers printed on thin polyethylene naphthalate (PEN) films. The top conductor layer is the resonator loaded with switches and biasing lines and the bottom conductor layer is the ground plane. Since the design is via-less, the two PEN films can be easily attached to transparent glasses. In addition, the stack provides high optical transparency with minimal visual impact, mechanical robustness due to the glass support, and compatibility with scalable additive manufacturing processes for low-cost, large-area integration.

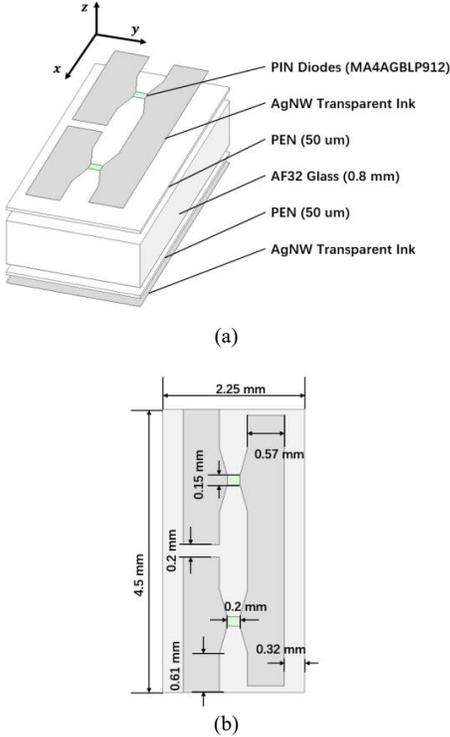

(a)

(b)

Fig. 1. The (a) layer stack and (b) top view of the proposed RIS unit cell.

On the top conductor layer in Fig. 1 (b), the unit cell consists of two H-shaped resonators that are serially connected along the x-axis, with biasing lines oriented perpendicular to the E-field polarization to increase the resonance capacitance and minimize interference with the electromagnetic response. A switch is placed at the center of each resonator to enable reconfigurability, while the bottom layer acts as a reflective ground plane. Based on this stack-up, the resonator is designed to achieve a 180° phase difference between the two switching states while maintaining high reflection magnitude. The H-shaped geometry is specifically chosen to provide sufficient current path variation between the ON and OFF states, thereby enabling effective phase control across the 5G mmWave band. The simulated reflection magnitude and phase are shown in Fig. 2. The unit cell is simulated in HFSS with periodic boundary conditions, where the switch (MA4AGBLP912) is modeled with a 5 Ω ON resistance and a 10 kΩ OFF resistance in parallel with a 30 fF junction capacitance. The AgNW conductor is represented as a 0.2 μm finite-conductivity sheet with conductivity of $5 \times 10^6$ S/m. The AF32 glass substrate has a permittivity of 5.1 and a loss tangent of 0.0084, while the PEN layer has a permittivity of 2.9 and a loss tangent of 0.003.

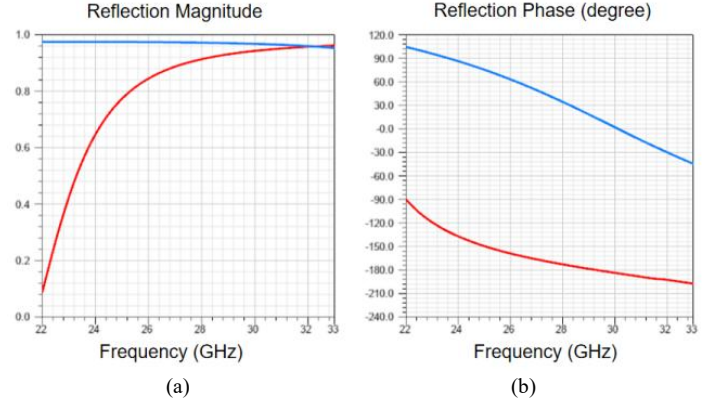

Fig. 2. Simulated reflection (a) magnitude and (b) phase.

Fig. 2 shows the simulated reflection magnitude and phase response of the unit cell under ON (red) and OFF (blue) states. As seen in the left plot, the OFF state maintains a nearly constant reflection magnitude above 0.95 across the 23.5–29.5 GHz band, while the ON state gradually increases from below 0.5 at 22 GHz to above 0.9 beyond 27 GHz. The right plot shows the corresponding reflection phase, where a clear separation between the two states is observed. The phase difference remains close to 180° within the targeted mmWave band, confirming the 1-bit reconfigurability of the proposed unit cell and supporting broadband operation. To understand the phase-shift mechanism, we plot the current distributions of the ON and OFF unit cells at 27.5 GHz in Fig. 3.

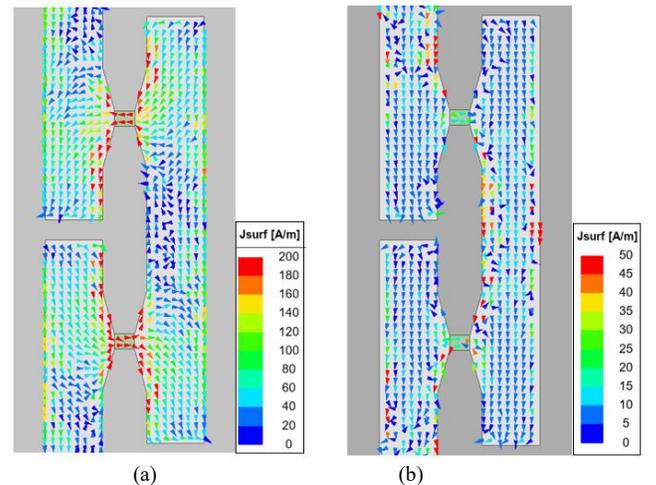

Fig. 3. The current distribution of the (a) ON and (b) OFF unit cell at 27.5 GHz.

As Fig. 3 (a) shows, the switches have the largest current concentration, connecting the vertical conductors. When the switches are OFF, the vertical conductors are not connected, which results in a shorter current path. Since the current path length is inversely proportional to the resonant frequency, the resonant frequency (the frequency where the reflection phase is 0°) of the unit cell increases from below our desired bandwidth to above the bandwidth, creating a phase difference in the 5G mmWave bands. To estimate its angular stability, more simulations with a variety of incident angles ranging from 0°, 15°, and 30° are shown in Fig. 4.

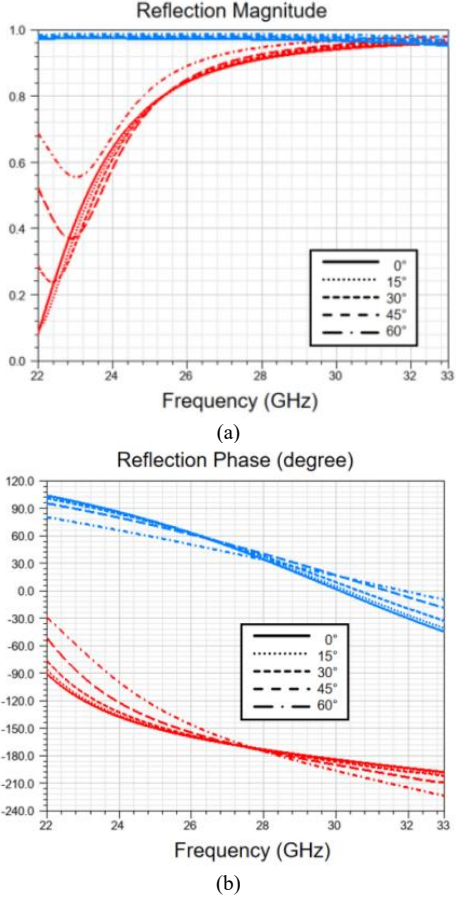

Fig. 4. The simulated reflection (a) magnitude, and (b) phase of the proposed unit cell at different states and incident angle, with red curves representing the ON states and blue curves OFF states.

From Fig. 4 (a), it can be seen that the ON reflection magnitude increases as the incident angle increases, while the OFF magnitude remains above 0.96. As the incident angle increases, the reflection magnitude is higher and the resonant frequency slightly increases from below 22 GHz to 23.5 GHz. The ON magnitude is above 50% after 23.5GHz for all incident angles. By comparing the OFF-reflection phase and the ON phase, we can see the phase difference in Fig. 4 (b). The unit cell is 1-bit, thus the design goal is to achieve a 180° phase difference with maximum reflection magnitude. As Fig. 4 (b) shows, the unit cell has a phase variation between 190° and 220° from 23.5 GHz to 29.5 GHz, with a maximum stable incident angle of up to 45°.

With the proposed unit cell, we further design the 9x20 RIS array, as depicted in Fig. 5 (a). Each column of the RIS has 9 serially connected resonators with PIN diodes, and the 20 columns are connected to the controller in Fig. 5 (b). The ESP32 microcontroller reads the commands from computer and assigns the ON/OFF patterns to each column of the RIS, with NMOS sinking current of 10 mA for each column.

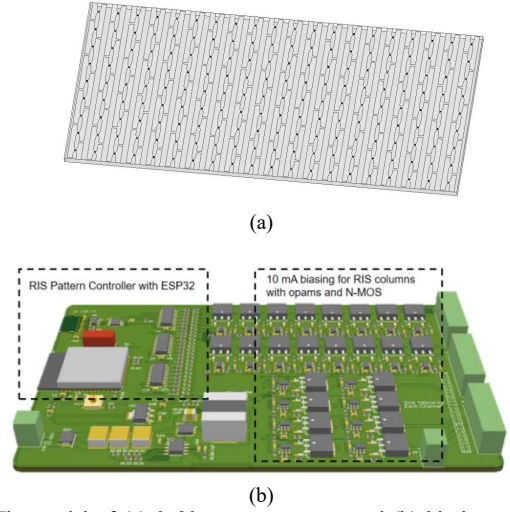

Fig. 5. The model of (a) 9x20 array prototype, and (b) biasing and control circuit.

## III. RIS FABRICATION AND MEASUREMENT

The large-area 9×20 RIS array was fabricated using an AgNWs-based ink formulation reported in our earlier studies [5], [12]. The RIS patterns were screen-printed on a PEN substrate using an AUREL 900 PA system with a stencil mesh count of 325, wire diameter of 20 μm, and emulsion thickness of 10 μm. Printing was performed at a speed of 220 mm/s to ensure uniform ink transfer. Following printing, the samples were dried in an oven at 80 °C for 5 minutes.

After washing and drying, the printed films underwent photonic sintering using a Novacentrix Pulse Forge 1300 system to achieve the desired conductivity. During sintering, the sample was placed 5 mm from the flash lamp and processed under optimized conditions: a pulse voltage of 375 V, pulse length of 700 μs, pulse rate of 0.5 Hz, and a total of 40 pulses. As shown in Fig. 6, the screen-printed AgNW conductor exhibits high optical transparency, allowing clear visibility through both the PEN substrate and the conductor. Finally, the conductor was mounted with PIN diodes and biasing wires to the control circuit.

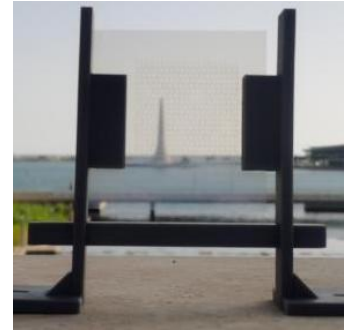

Fig. 6. The printed AgNW conductor on the PEN substrate

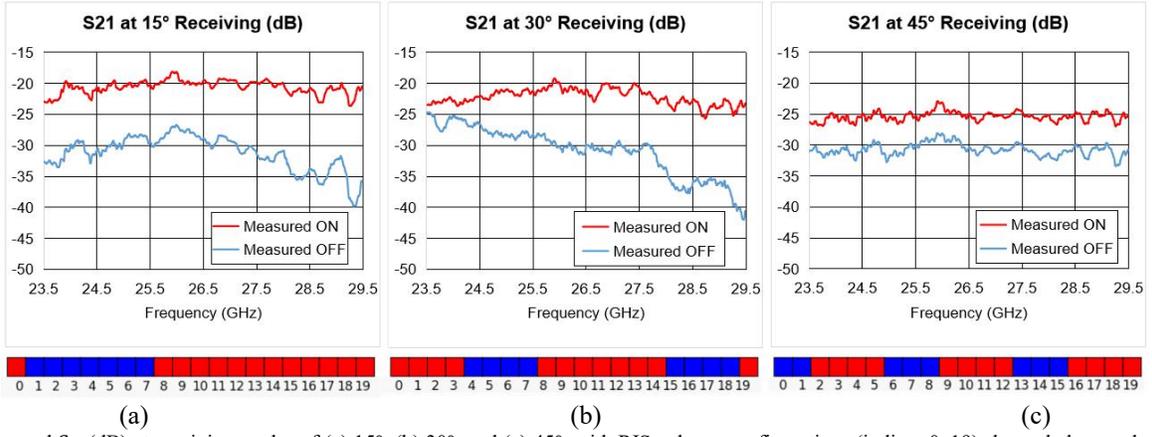

Fig. 8. Measured $S_{21}$ (dB) at receiving angles of (a) 15°, (b) 30°, and (c) 45°, with RIS column configurations (indices 0–19) shown below each plot, where red and blue indicate ON and OFF states, respectively.

To evaluate the RIS performance, a pair of 20 dBi WR34 antennas are placed in a secondary-line-of-sight link, as Fig. 7 shows. Both antennas are positioned 24 cm away from the RIS, which provides a good trade-off between tapering and spillover. The transmitting antenna is set at 0° incidence along the -Z axis, while the receiving antenna is at 15°, 30°, 45° on the Z-O-Y plane. The antennas are connected to a VNA, to measure from 23.5 GHz to 29.5 GHz, and for each receiving angle we compare the $S_{21}$ with the RIS pattern ON and all OFF calculated from the following equation (1).

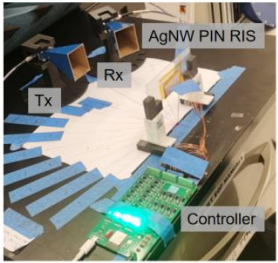

Fig. 7. The measurement setup of the RIS.

$$\varphi_i = f(\boldsymbol{k}_{incidence} \cdot \boldsymbol{r}_i - \boldsymbol{k}_{reflect} \cdot \boldsymbol{r}_i) \quad (1)$$

In (1), $\varphi_i$ is the reflection phase of the column i at its center position. $\boldsymbol{k}_{incidence}$ is the incident wave vector, and $\boldsymbol{r}_i$ is the center position of the column $i$. $\boldsymbol{k}_{reflect}$ is the desired reflection wave vector, which is along +z in this setup. $f$ is the 1-bit truncation function shown in equation (2).

$$f(\varphi) = \begin{cases} \arg(\Gamma_{ON}), & |e^{j\varphi} - \Gamma_{ON}| < |e^{j\varphi} - \Gamma_{OFF}| \\ \arg(\Gamma_{OFF}), & |e^{j\varphi} - \Gamma_{ON}| \geq |e^{j\varphi} - \Gamma_{OFF}| \end{cases} \quad (2)$$

In (2), $\Gamma_{ON}$ is the reflection coefficient of the ON unit cell, and $\Gamma_{OFF}$ is that of the OFF unit cell. The truncation takes the state in which the distance between the ideal reflection coefficient and the state reflection is closer.

The measured gain enhancement at 15° receiving angle is shown in Fig. 8 (a). We can see that from 23.5 GHz to 29.5 GHz, the measured enhancement ranges from 10 dB to 15 dB

As the incident angle increases from 15° to 45°, we can see the decrease of gain enhancement from 10 ∼ 15 dB in Fig. 8 (a) and (b) to 5 dB in Fig. 8 (c). This is because of the variation in the reflection phase in Fig. 4 (b), as well as the fabrication error in printing and film attachment. From the measurement results, the RIS shows a broad band 10 ∼ 15 dB signal enhancement from 23.5 GHz to 29.5 GHz with maximum 30° incidence.

TABLE I
LITERATURE COMPARISON

| | Transparency (%) | Bandwidth (GHz) | Gain Enhancement (dB) | Array Size | Material |
|---|---|---|---|---|---|
| **This work** | **80 +** | **23.9 ∼ 29.5** | **10 ∼ 15** | **9 x 20** | **AgNW** |
| [6] | 49.3 | 3.36 ∼ 3.55 | 12 | 12 x 10 | Metal Mesh |
| [7] | 79 | 3.53 ∼ 3.7 | NA | 16 x 8 | Metal Mesh |
| [14] | 60+ | 26.75 ∼ 28.25 | 12+ | 10 x 10 | Metal Mesh |

Comparing our results with published papers in TABLE I, we can see that our proposed AgNW RIS provides the largest bandwidth in the 5G mmWave range while maintaining high optical transparency above 80%. In contrast to metal-mesh based designs, which either operate at sub-6 GHz or within narrower mmWave windows, our approach achieves broadband operation with consistent phase control. Moreover, the demonstrated 10 ∼ 15 dB gain enhancement across 0° ∼ 30° angles confirms robust reconfigurability. The use of screen-printable AgNW conductors enables a via-less, low-cost, and scalable fabrication route, offering practical advantages for large-area integration on transparent surfaces such as windows and displays.

## IV. CONCLUSION

In conclusion, we have designed, fabricated and tested a 9x20 RIS screen-printed with optically transparent AgNW ink and mounted with PIN diodes. The RIS shows good optical transparency and can enhance 5 ∼ 15 dB of the signal power in the secondary-line-of-sight measurement setup, with a broad bandwidth from 23.5 GHz to 29.5 GHz.